\documentclass[aps,prl,twocolumn,superscriptaddress,showpacs,preprintnumbers,amsmath,amssymb, nofootinbib]{revtex4-2}
\UseRawInputEncoding

\usepackage{graphicx}
\usepackage{dcolumn}
\usepackage{lineno}
\usepackage{color}
\usepackage[dvipsnames]{xcolor}
\usepackage{comment}
\usepackage{multirow}
\usepackage{amssymb}
\usepackage{url}
\usepackage{hyperref}
\usepackage{verbatim}
\usepackage{orcidlink}
\usepackage{xspace}
\usepackage{float}
\usepackage{footnote}
\restylefloat{table}
\usepackage{caption}
\def\babar{\mbox{\slshape B\kern-0.3em{\smaller[2] A}\kern-0.1em B\kern-0.3em{\smaller[2] A\kern-0.2em R}} }

\begin{document}
\newcommand\y[1]{\Upsilon(#1S)}
\newcommand\Cb{\chi_{b[0,1,2]}(1P)}
\newcommand\cb[1]{\chi_{b#1}(1P)}
\newcommand\bn{B^0}
\newcommand\bc{B^+}
\newcommand\x{X(3872)}
\newcommand\X{X(3915)}
\newcommand\jp{J/\psi}
\newcommand\ks{K^{0}_{S}}

\newcommand\bll{\mt B(V\rightarrow\ell_1\ell_2)}
\newcommand\bee{\mt B(V\rightarrow e^+ e^-)}
\newcommand\ppee{\Upsilon(2S)\rightarrow\pi^+\pi^-\Upsilon(1S)[\rightarrow e^+e^-]}
\newcommand\ppmm{\Upsilon(2S)\rightarrow\pi^+\pi^-\Upsilon(1S)[\rightarrow \mu^+\mu^-]}
\newcommand\pptt{\Upsilon(2S)\rightarrow\pi^+\pi^-\Upsilon(1S)[\rightarrow \tau^+\tau^-]}
\newcommand\ppem{\Upsilon(2S)\rightarrow\pi^+\pi^-\Upsilon(1S)[\rightarrow e^\pm\mu^\mp]}
\newcommand\ppmt{\Upsilon(2S)\rightarrow\pi^+\pi^-\Upsilon(1S)[\rightarrow \mu^\pm\tau^\mp]}
\newcommand\ppet{\Upsilon(2S)\rightarrow\pi^+\pi^-\Upsilon(1S)[\rightarrow e^\pm\tau^\mp]}
\newcommand\ppgem{\Upsilon(2S)\rightarrow\pi^+\pi^-\Upsilon(1S)[\rightarrow\gamma e^\pm\mu^\mp]}
\newcommand\ppgmt{\Upsilon(2S)\rightarrow\pi^+\pi^-\Upsilon(1S)[\rightarrow\gamma \mu^\pm\tau^\mp]}
\newcommand\ppget{\Upsilon(2S)\rightarrow\pi^+\pi^-\Upsilon(1S)[\rightarrow\gamma e^\pm\tau^\mp]}
\newcommand\GGEE{\gamma\gamma e^+ e^-}
\newcommand\GGMM{\gamma\gamma \mu^+ \mu^-}
\newcommand\GGLL{\gamma\gamma \ell^-\ell^+}
\newcommand\GEE{\gamma e^+ e^-}
\newcommand\GMM{\gamma \mu^+ \mu^-}
\newcommand\GEM{\gamma e^\pm \mu^\mp}
\newcommand\GET{\gamma e^\pm \tau^\mp}
\newcommand\GMT{\gamma \mu^\pm \tau^\mp}
\newcommand\EM{e^\pm \mu^\mp}
\newcommand\ET{e^\pm \tau^\mp}
\newcommand\MT{\mu^\pm \tau^\mp}
\newcommand\EP{e^+e^-}
\newcommand\GLL{\gamma \ell^+\ell^-}
\newcommand\LL{\ell_1^\pm\ell_2^\mp}

\newcommand\ppy{\y2\rightarrow\pi^+\pi^-\y1}
\newcommand\pPy{\y2\rightarrow\pi^0\pi^0\y1}
\newcommand\yee{\y1\rightarrow e^+e^-}
\newcommand\ymm{\y1\rightarrow\mu^+\mu^-}
\newcommand\ytt{\y1\rightarrow\tau^+\tau^-}
\newcommand\Yee{\y2\rightarrow e^+e^-}
\newcommand\Ymm{\y2\rightarrow\mu^+\mu^-}
\newcommand\yll{\y1\rightarrow\ell_2^\pm\ell^{\mp}_2}
\newcommand\Yll{\y1\rightarrow\ell^+\ell^-}
\newcommand\ygll{\y1\rightarrow\gamma\ell_1^\pm\ell^{\mp}_2}
\newcommand\yem{\y1\rightarrow e^\pm\mu^\mp}
\newcommand\ymt{\y1\rightarrow \mu^\pm\tau^\mp}
\newcommand\yet{\y1\rightarrow e^\pm\tau^\mp}
\newcommand\ylt{\y1\rightarrow\ell^\pm\tau^\mp}
\newcommand\ygem{\y1\rightarrow \gamma e^\pm\mu^\mp}
\newcommand\ygmt{\y1\rightarrow \gamma \mu^\pm\tau^\mp}
\newcommand\yget{\y1\rightarrow \gamma e^\pm\tau^\mp}
\newcommand\yglt{\y1\rightarrow \gamma \ell^\pm\tau^\mp}
\newcommand\te{\tau^-\rightarrow e^-\bar{\nu}_e\nu_\tau}
\newcommand\tm{\tau^-\rightarrow \mu^-\bar{\nu}_\mu\nu_\tau}
\newcommand\tp{\tau^-\rightarrow\pi^-\nu_\tau}
\newcommand\tP{\tau^-\rightarrow\pi^-\pi^+\pi^-\nu_\tau}

\newcommand\ggy[1]{\Upsilon(2S)\rightarrow\gamma\chi_{b#1}(1P)[\rightarrow\gamma\y1]}
\newcommand\gc[1]{\y2\rightarrow\gamma\chi_{b#1}(1P)}
\newcommand\gC{\y2\rightarrow\gamma\Cb}
\newcommand\cgyee[1]{\chi_{b#1}(1P)\rightarrow\gamma\y1 [\rightarrow e^+e^-]}
\newcommand\cgee[1]{\chi_{b#1}(1P)\rightarrow\gamma e^+e^-}
\newcommand\cgmm[1]{\chi_{b#1}(1P)\rightarrow\gamma\mu^+\mu^-}
\newcommand\cgll[1]{\chi_{b#1}(1P)\rightarrow\gamma\ell^+\ell^-}
\newcommand\cgymm[1]{\chi_{b#1}(1P)\rightarrow\gamma\y1 [\rightarrow\mu^+\mu^-]}
\newcommand\Cgyee{\Cb\rightarrow\gamma\y1 [\rightarrow e^+e^-]}
\newcommand\Cgy{\Cb\rightarrow\gamma\y1}
\newcommand\cgy[1]{\chi_{b#1}(1P)\rightarrow\gamma\y1}
\newcommand\Cgymm{\Cb\rightarrow\gamma\y1 [\rightarrow\mu^+\mu^-]}
\newcommand\cem[1]{\chi_{b#1}(1P)\rightarrow e^\pm \mu^\mp}
\newcommand\cmt[1]{\chi_{b#1}(1P)\rightarrow\mu^\pm\tau^\mp}
\newcommand\cet[1]{\chi_{b#1}(1P)\rightarrow e^\pm\tau^\mp}
\newcommand\Cem{\Cb\rightarrow e^\pm\mu^\mp}
\newcommand\Cmt{\Cb\rightarrow \mu^\pm\tau^\mp}
\newcommand\Cet{\Cb\rightarrow e^\pm\tau^\mp}
\newcommand\Clt{\Cb\rightarrow \ell^\pm\tau^\mp}
\newcommand\clt[1]{\chi_{b#1}(1P)\rightarrow\ell^\pm\tau^\mp}
\newcommand\Cll{\Cb\rightarrow \ell_1^\pm\ell^{\mp}_2}
\newcommand\cll[1]{\chi_{b#1}(1P)\rightarrow \ell_1^+\ell^-_2}
\newcommand\gCll{\Upsilon(2S)\rightarrow\gamma\chi_{bJ}[\rightarrow \ell_1^\pm\ell^{\mp}_2]}

\newcommand\jok{B\rightarrow\jp\omega K}
\newcommand\jokp{B^{+}\rightarrow\jp\omega K^{+}}
\newcommand\joks{B^{0}\rightarrow\jp\omega K^{0}_{S}}
\newcommand\jx{B\rightarrow\jp X}
\newcommand\xjo{\x\rightarrow\jp\omega}
\newcommand\Xjo{\X\rightarrow\jp\omega}
\newcommand\xk{B\rightarrow\x K}
\newcommand\Xk{B\rightarrow\X K}
\newcommand\xkp{B^{+}\rightarrow\x K^{+}}
\newcommand\xks{B^{0}\rightarrow\x K^{0}_{S}}
\newcommand\Xkp{B^{+}\rightarrow\X K^{+}}
\newcommand\Xks{B^{0}\rightarrow\X K^{0}_{S}}
\newcommand\jll{\jp\rightarrow\ell\ell}
\newcommand\oP{\omega\rightarrow\pi^+\pi^0\pi^-}
\newcommand\pgg{\pi^0\rightarrow\gamma\gamma}
\newcommand\kpp{\ks\rightarrow\pi^+\pi^-}


\newcommand{\tev}{\ensuremath{\mathrm{\,Te\kern -0.1em V}}\xspace}
\newcommand{\gev}{\ensuremath{\mathrm{\,Ge\kern -0.1em V}}\xspace}
\newcommand{\mev}{\ensuremath{\mathrm{\,Me\kern -0.1em V}}\xspace}
\newcommand{\kev}{\ensuremath{\mathrm{\,ke\kern -0.1em V}}\xspace}
\newcommand{\ev}{\ensuremath{\mathrm{\,e\kern -0.1em V}}\xspace}
\newcommand{\gevc}{\ensuremath{{\mathrm{\,Ge\kern -0.1em V\!/}c}}\xspace}
\newcommand{\mevc}{\ensuremath{{\mathrm{\,Me\kern -0.1em V\!/}c}}\xspace}
\newcommand{\gevcc}{\ensuremath{{\mathrm{\,Ge\kern -0.1em V\!/}c^2}}\xspace}
\newcommand{\mevcc}{\ensuremath{{\mathrm{\,Me\kern -0.1em V\!/}c^2}}\xspace}

\def\inch   {\ensuremath{\rm \,in}\xspace} 
\def\ft   {\ensuremath{\rm \,ft}\xspace}
\def\km   {\ensuremath{{\rm \,km}}\xspace}
\def\cm   {\ensuremath{{\rm \,cm}}\xspace}
\def\cma  {\ensuremath{{\rm \,cm}^2}\xspace}
\def\mm   {\ensuremath{{\rm \,mm}}\xspace}
\def\mma  {\ensuremath{{\rm \,mm}^2}\xspace}
\def\mum  {\ensuremath{{\,\mu\rm m}}\xspace}
\def\muma       {\ensuremath{{\,\mu\rm m}^2}\xspace}
\def\nm   {\ensuremath{{\rm \,nm}}\xspace}
\def\fm   {\ensuremath{{\rm \,fm}}\xspace}
\def\nm         {\ensuremath{{\rm \,nm}}\xspace}   
\def\mrad{\ensuremath{\rm \,mrad}\xspace}               
\newcommand\e[1]{10^{-#1}}
\newcommand\mt[1]{\mathcal{#1}}
\newcommand\qq[1]{{#1}\bar{#1}}
\newcommand\ecm{E_{\rm cm}}
\newcommand\mbc{M_{\rm bc}}
\newcommand\de{\Delta E}
\newcommand\dm{\Delta M}
\newcommand\mr[1]{ M^{{\rm recoil}}_{\rm #1}}
\newcommand\T[1]{\cos(\theta_{#1})}
\newcommand\fb{{\rm fb}^{-1}}
\newcommand\mrgl{\mr{\gamma_1\ell_1}}
\newcommand\mrg{\mr{\gamma_1}}
\newcommand\mrgm{\mr{\gamma_1\mu}}
\newcommand\mrge{\mr{\gamma_1 e}}
\newcommand\mrpp{\mr{\pi\pi}}
\newcommand\mrppm{\mr{\pi\pi\mu}}
\newcommand\mrppe{\mr{\pi\pi e}}
\newcommand\mrppl{\mr{\pi\pi\ell}}
\newcommand\mrppgm{\mr{\pi\pi\mu\gamma}}
\newcommand\mrppge{\mr{\pi\pi e\gamma}}
\newcommand\mrppgl{\mr{\pi\pi\ell\gamma}}
\newcommand\erpp{E^{\rm recoil}_{\pi\pi}}
\newcommand\pvt{p^{\tau}_{\rm vis}}
\newcommand\N[1]{N_{\rm #1}}
\newcommand\Nm{\N{\mu}}
\newcommand\Ne{\N{e}}
\newcommand\nbb{N_{\rm B\bar{B}}}
\newcommand\ndm{\delta M}
\newcommand\mrgg{\mr{\gamma_1\gamma_2}}
\newcommand\cospp{\T{\pi\pi}}
\newcommand\cosgc{\T{\gamma\chi}}
\newcommand\m[1]{M_{\rm #1}}
\newcommand\E[1]{E_{\rm #1}}
\newcommand\p[1]{|\vec{p}_{\rm #1}|}
\newcommand\chifc{\chi^2_{\rm 4C}}
\newcommand\mjo{\m{\jp\omega}}
\newcommand\fom{F_{\rm OM}}

\newcommand\CVL{C_{VL}^{q\ell_1\ell_2}}
\newcommand\CVR{C_{VR}^{q\ell_1\ell_2}}
\newcommand\CDL{C_{DL}^{q\ell_1\ell_2}}
\newcommand\CDR{C_{DR}^{q\ell_1\ell_2}}
\newcommand\CAL{C_{AL}^{q\ell_1\ell_2}}
\newcommand\CAR{C_{AR}^{q\ell_1\ell_2}}
\newcommand\CTR{C_{TR}^{q\ell_1\ell_2}}
\newcommand\CTL{C_{TL}^{q\ell_1\ell_2}}
\newcommand\CSL{C_{SL}^{q\ell_1\ell_2}}
\newcommand\CSR{C_{SR}^{q\ell_1\ell_2}}
\newcommand\tcr[1]{\textcolor{red}{#1}}

\newcommand\Wq[1]{C_{#1}^{q\ell\ell^\prime}}
\newcommand\W[1]{C_{#1}^{\ell\ell^\prime}}

\newcommand{\pyhf}{\texttt{pyhf}}

\include{reference}

\onecolumngrid

\vspace*{-3\baselineskip}
\resizebox{!}{3cm}{\includegraphics{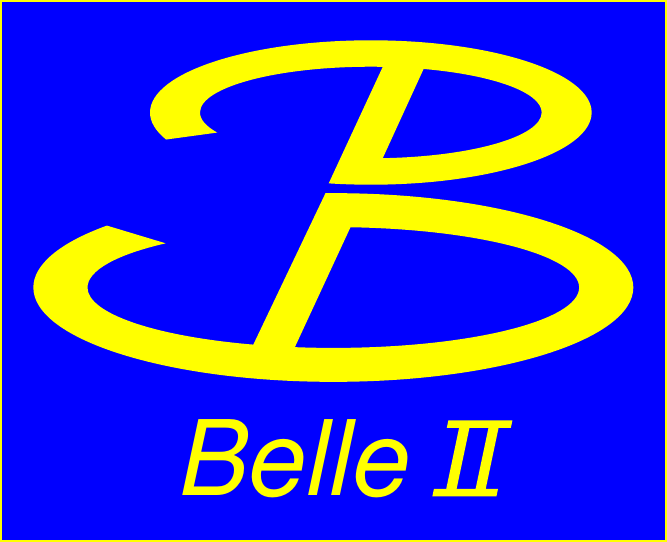}}

\preprint{Belle~II Preprint 2026-005}
\preprint{KEK Preprint 2026-1}

\vspace*{-2\baselineskip}
\title{ \quad\\[0.5cm] Searches for charged-lepton-flavor violation in $\chi_{bJ}(1P)$ decays}

  \author{M.~Abumusabh\,\orcidlink{0009-0004-1031-5425}} 
  \author{I.~Adachi\,\orcidlink{0000-0003-2287-0173}} 
  \author{A.~Aggarwal\,\orcidlink{0000-0002-5623-3896}} 
  \author{L.~Aggarwal\,\orcidlink{0000-0002-0909-7537}} 
  \author{H.~Ahmed\,\orcidlink{0000-0003-3976-7498}} 
  \author{Y.~Ahn\,\orcidlink{0000-0001-6820-0576}} 
  \author{H.~Aihara\,\orcidlink{0000-0002-1907-5964}} 
  \author{N.~Akopov\,\orcidlink{0000-0002-4425-2096}} 
  \author{S.~Alghamdi\,\orcidlink{0000-0001-7609-112X}} 
  \author{M.~Alhakami\,\orcidlink{0000-0002-2234-8628}} 
% ...
\collaboration{The Belle and Belle II Collaborations}

\begin{abstract}
  We report the first searches for charged-lepton-flavor violation in decays of $\chi_{bJ}(1P)$ ($J=0, 1,$ and $2$) to a pair of charged leptons
  using 158 million $\Upsilon(2S)$ decays collected with the Belle detector in $e^+e^-$ collisions at the KEKB collider.
  No significant signal is observed, and we set upper limits on the branching fractions for $\chi_{bJ}(1P)$ decays
  to $e^\pm\mu^\mp$ at the level of $10^{-6}$ and to $e^\pm\tau^\mp$ or $\mu^\pm\tau^\mp$ at the level of $10^{-5}$.
  Limits on $\chi_{b0}(1P)$ decays are translated into bounds on the corresponding Wilson coefficients of scalar operators that mediate charged-lepton-flavor violation.
\end{abstract}


\maketitle

Charged-lepton-flavor violation (CLFV) is an unambiguous signature of physics beyond the standard model (SM)~\cite{Raidal:2008jk,deGouvea:2013zba,Celis:2014asa}.
Extensive searches for CLFV in decays of the $\tau$ lepton~\cite{BaBar:2009hkt,Belle:2021ysv},
mesons~\cite{Alonso:2015sja,Abada:2015zea}, and the Higgs boson~\cite{CMS:2021rsq,ATLAS:2023mvd} have placed stringent limits on new interactions.
CLFV has so far been searched for in decays of pseudoscalar mesons like $\pi^0$~\cite{NA62:2021zxl} and $\eta$~\cite{White:1995jc} and vector quarkonia such as $J/\psi$~\cite{BESIII:2021slj} and $\Upsilon(nS)$~\cite{BaBar:2010vxb,BaBar:2021loj,Belle:2022cce,Dhamija:2024}.

In this Letter, we report the first searches for CLFV in $\chi_{bJ}(1P)$ decays with $J=0, 1,$ and $2$.
The $\chi_{b0}(1P)$ decays provide a unique low-energy probe of CLFV in the scalar sector, complementary to direct searches for CLFV in Higgs boson decays.
Similarly, the $\chi_{b1}(1P)$ and $\chi_{b2}(1P)$ decays provide the first searches for CLFV in decays of axial-vector and tensor particles, respectively~\cite{Hazard:2016fnc}.
Although not expected within the SM, interactions beyond the SM can induce
$\cll{J}$ with $\ell_1\neq\ell_2$ and $\ell_1,\ell_2=e,\mu,\tau$~\footnote{Charge-conjugate decay modes are implied throughout.}
at experimentally observable rates~\cite{Hazard:2016fnc}, providing strong motivation for these searches.
The observation of any such decay would clearly indicate the presence of new particles and interactions beyond the SM.

The $\chi_{bJ}(1P)$ states are copiously produced via $\Upsilon(2S)\to \gamma\chi_{bJ}(1P)$ in $e^+e^-$ collisions.
This study uses a sample of $(158 \pm 4) \times 10^6$ $\y2$ mesons~\cite{Wang:2011ny2}, the largest collected to date, produced in 24.7~$\fb$ of data recorded by the Belle detector~\cite{Belle:2000cnh, Belle:2012iwr}
at a center-of-mass energy of 10.023\gev at the KEKB collider~\cite{Kurokawa:2001nw,Abe:2013fma}.

Several decays are simulated to estimate reconstruction efficiencies and background contributions.
We simulate $\y2$ decays using EVTGEN~\cite{Lange:2001uf},
leptonic $\cb{J}$ decays using the multi-body phase-space decay model,
radiative decays using the helicity-amplitude decay model,
and leptonic $\y1$ decays using the vector-meson-to-dilepton decay model.
Processes such as $e^+e^- \to \ell^+\ell^-$, $q\bar{q}~(q=u,d,s,c)$, $\gamma\gamma$, and $\gamma q\bar{q}$ 
that potentially contribute to backgrounds are simulated using PYTHIA~6.4~\cite{Sjöstrand_2006}.
We use PHOTOS~\cite{Barberio:1993qi} to simulate final-state radiation and GEANT3~\cite{Brun:1987ma} to model the response of the Belle detector.

To characterize background rates, estimate systematic uncertainties, and validate our analysis procedure,
we study cascade decays of $\cgy{J}$ with subsequent $\yee$ or $\ymm$ decays as our control modes.
These control modes are chosen because direct decays of $\chi_{bJ}(1P)$ to same-flavor lepton pairs proceed via two photons or the Higgs boson in the SM,
and have branching fractions of ${\mathcal{O}}(10^{-20})\text{--}{\mathcal{O}}(10^{-9})$~\cite{Godfrey:2015vda}, making them unmeasurable.
In the searches for $\chi_{bJ}(1P) \to e^\pm \tau^\mp$ or $\mu^\pm \tau^\mp$, we reconstruct the $\tm$ and $\te$ decay modes, respectively,
to minimize backgrounds from $e^+ e^- \to \ell^+ \ell^-$ events, where $\ell = e, \mu$.

We select events with exactly two oppositely charged tracks, one identified as an electron and the other as a muon.
Each track is required to have a momentum greater than 200\mev~\footnote{Natural units ($\hbar=c=1$) are used throughout this letter, and all variables are defined in the laboratory frame, unless otherwise mentioned.}
and a distance of closest approach to the $e^+e^-$ collision point of less than 4.5\cm along the $e^-$ beam direction and 1.5\cm in the plane transverse to it.
A track is identified as an electron if its electron likelihood ratio exceeds 0.5; otherwise, it is identified as a muon when its muon likelihood ratio exceeds 0.9.
The efficiencies of this selection are 92.4\% and 93.8\% for electrons and muons, respectively~\cite{Hanagaki:2001fz,Abashian:2002bd}.
We recover the energy that an electron may lose to bremsstrahlung by adding to its four-momentum the four-momenta of photon candidates found within $60~{\rm{mrad}}$ of its momentum.
To make the datasets mutually exclusive, we require the momenta of electrons or muons from $\cb{J}$ decays exceed 3.5\gev and that from $\tau$ decays fall below 3.5\gev.

All photons must have energies above 50\mev\ to reduce the contamination from beam background.
The energies of the photons from $\Upsilon(2S)\to\gamma\chi_{bJ}(1P)$ and $\cgy{J}$ decays, labeled $E_{{\gamma}_1}$ and $E_{{\gamma}_2}$, respectively,
are known to mostly satisfy the relation $E_{{\gamma}_1} <  E_{{\gamma}_2}$~\cite{CLEO:2010xuh, BaBar:2014och}.
To reduce combinatorial background, we require $E_{{\gamma}_1}$ $<$ 250\mev and $E_{{\gamma}_2}$ $>$ 250\mev.

For the $\cb{J}\to e^\pm\mu^\mp$ decays and the control modes, a four-constraint kinematic fit is performed
using energy-momentum conservation and the measured uncertainties of the final-state momenta.
Kinematically inconsistent events are suppressed by a loose requirement on the fit $\chi^2$ value, $\chi^2_{\rm 4C} < 1000$, a selection that retains nearly all signal decays.

The control mode $\chi_{bJ}(1P)\to\gamma e^+e^-$ is mimicked by the radiative Bhabha process, 
whose cross section is significantly larger than that for $\Upsilon(2S)$ production. 
A veto that rejected 99\% of Bhabha events was applied during data taking~\cite{CHEON2002548}. 
To further suppress such events, we remove regions dominated by Bhabha backgrounds, 
characterized by $\theta_{e^-}<0.75~\mathrm{rad}$ and $\theta_{\gamma_1}<0.6~\mathrm{rad}$, or $\theta_{\gamma_1}>2.2~\mathrm{rad}$, 
where $\theta_{e^-}$ and $\theta_{\gamma_1}$ are the angles of the final-state electron and the $\gamma_1$ candidate with respect to the electron beam.
Additionally, we reject any event with an $e^-$ track with momentum above $6.4\mathrm{GeV}$. 
We estimate the efficiency of the control mode using events that pass a dedicated simulation of the Bhabha veto.

In terms of the total energy $E_{X}$ and momentum ${\vec{p}}_{X}$ of a set of particles ${X}$,
the mass recoiling against that set is given by
\begin{equation}
M^{\rm{recoil}}_{X} = \sqrt{(E_{e^+e^-}-E_{X})^2-({\vec{p}}_{e^+e^-}-{\vec{p}}_{X})^2},
\label{eq:mrecoil}
\end{equation}
where $E_{e^+e^-}$ and ${\vec{p}}_{e^+e^-}$ are the total energy and momentum of the colliding electron and positron.
Control events are required to have $\mrgg$ in [9.43, 9.51]\gev, encompassing the $\y1$ resonance.
Both control and signal events are required to have $\mrg$ in [9.81, 9.95]\gev, covering the $\cb{J}$ states.
Also, we require $M^{\rm{recoil}}_{\gamma_1\ell_1}$ to be in [0.6, 2.8]\gev, which extends approximately 2.5 times the mass resolution in either direction of the tau mass, for $\chi_{bJ}(1P)\to\ell_1^\pm\tau^\mp$ searches, where $\ell_1$ refers to the higher-energy lepton in the final state.

In 40\% of events, multiple $\y2$ candidates are formed due to photons from beam background and final-state radiation.
For $\cem J$ and the control modes, we retain the candidate with the lowest $\chifc$, which selects the true candidate 98\% of the time in simulation.
For $\chi_{bJ}(1P)\to \ell_1^\pm \tau^\mp$, we retain the candidate whose $\gamma_1$ energy yields an $M^{\rm recoil}_{\gamma_1\ell_1}$ value closest to the known $\tau$ mass~\cite{Belle-II:2023izd}, selecting the true candidate 75\% of the time. In simulated events, best-candidate selection does not bias the analysis results.

\begin{figure}[hbtp]
  \begin{center}
  \includegraphics[width=0.9\linewidth]{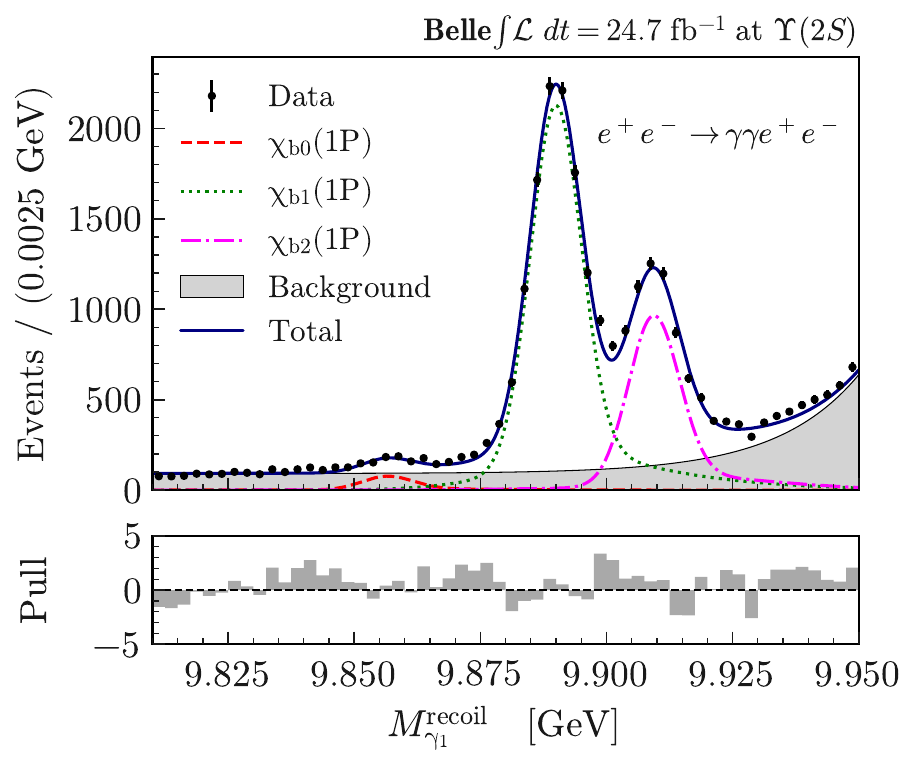}
  \includegraphics[width=0.9\linewidth]{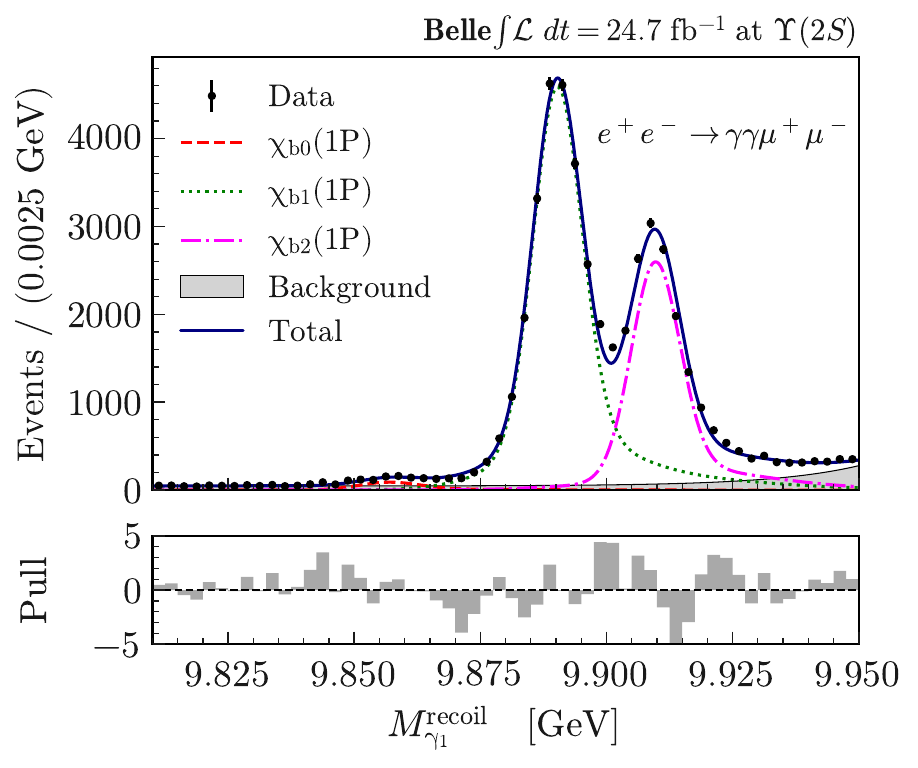}
  \end{center}
  \caption{$\mrg$ distributions for $\EP\to\GGEE$ (top) and $\EP\to\GGMM$ (bottom) overlaid with fit results.}
  \label{fig:ll}
  \end{figure}

To obtain the yields of the control modes $N_{J, \ell\ell}$, we perform unbinned maximum-likelihood fits to $\mrg$ distributions separately for $\GGEE$ and $\GGMM$ candidates.
For both, we model the probability density function (PDF) for each $\cb{J}$ as a sum of two Gaussian functions and a bifurcated Gaussian function, all sharing a common central value.
The central value and the widths of the PDF for $\cb1$ are determined by the fit to data, while the parameters for the $\cb0$ and $\cb2$ components are fixed with respect to them.
We model each background distribution as the sum of an exponential PDF and a constant PDF. Figure~\ref{fig:ll} shows the distributions of $\mrg$ with the fitted PDF components overlaid.
From each $N_{J, \ell\ell}$, we calculate the product of branching fractions for $\gc{J}$, $\cgy{J}$, and $\Yll$ as
\begin{equation}
\mt{B}(\y2\to\gamma\gamma\ell^+\ell^-) = \frac{N_{J,\ell\ell}}{N_{\y2}\epsilon_{J,\ell\ell}},
\label{equ:bf}
\end{equation}
where $N_{\y2}$ is the number of $\y2$ mesons in the data, and $\epsilon_{J, \ell\ell}$ is the efficiency to reconstruct $\y2\to\gamma\gamma\ell^+\ell^-$ via $\chi_{bJ}(1P)$.
The efficiencies for $\GGEE$ are 16\%, 16\%, and 13\% for $J = 0, 1,$ and $2$, respectively; for $\GGMM$ they are 28\%, 30\%, and 29\%, respectively.
Table~\ref{tab:control} summarizes the results, which agree with the products of the world averages of the individual branching fractions
in $\y2\to\gamma\gamma\ell^+\ell^-$~\cite{ParticleDataGroup:2024cfk}.

\begin{table}[hbtp]
\begin{center}
\begin{tabular}{l|c|c}
\hline\hline 
Decay & $\y2\to\GGEE$   & $\y2\to\GGMM$   \\\hline
$J=0$ & $0.175\pm0.018$ & $0.147\pm0.010$ \\
$J=1$ & $5.179\pm0.056$ & $5.912\pm0.041$ \\
$J=2$ & $2.799\pm0.058$ & $3.267\pm0.037$ \\ \hline 
\hline
\end{tabular}
\end{center}
\caption{Products of branching fractions in units of $\e{4}$ for $\gc{J}$, $\cgy{J}$ and $\Yll$ measured in the control modes. Uncertainties are statistical only.}
\label{tab:control}
\end{table}

\begin{figure}[hbtp!]
  \begin{center}
  \includegraphics[width=0.9\linewidth]{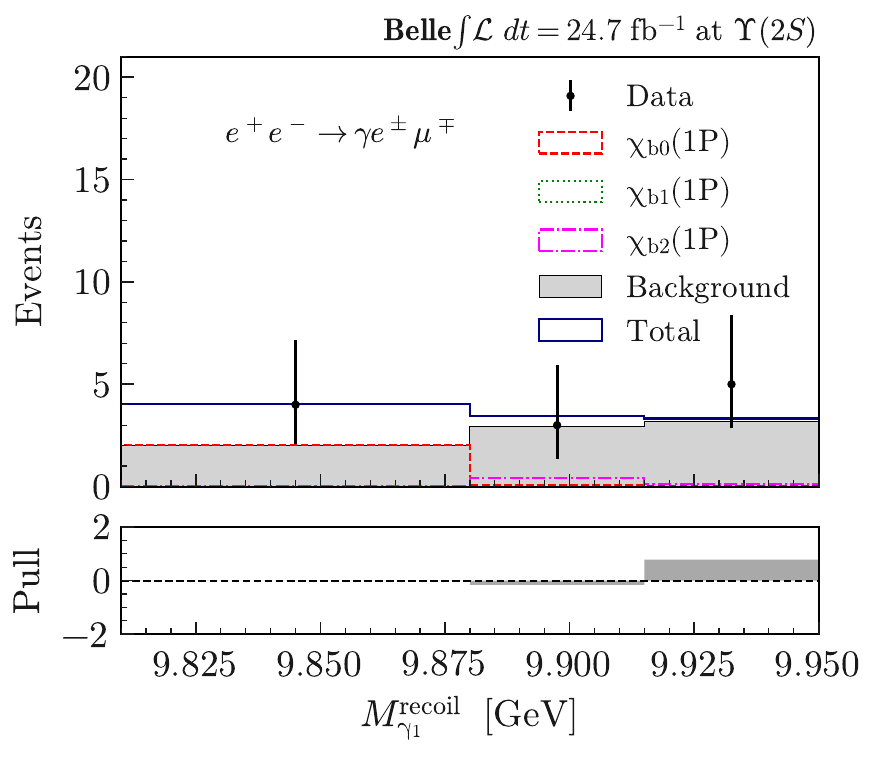}
  \end{center}
  \caption{$\mrg$ distribution for $\EP\to\GEM$ candidates. The $\chi_{b1}(1P)$ component is too small to be visible.}
  \label{fig:em}
  \end{figure}

\begin{figure*}[hbtp]
\begin{center}
\includegraphics[width=0.4\textwidth]{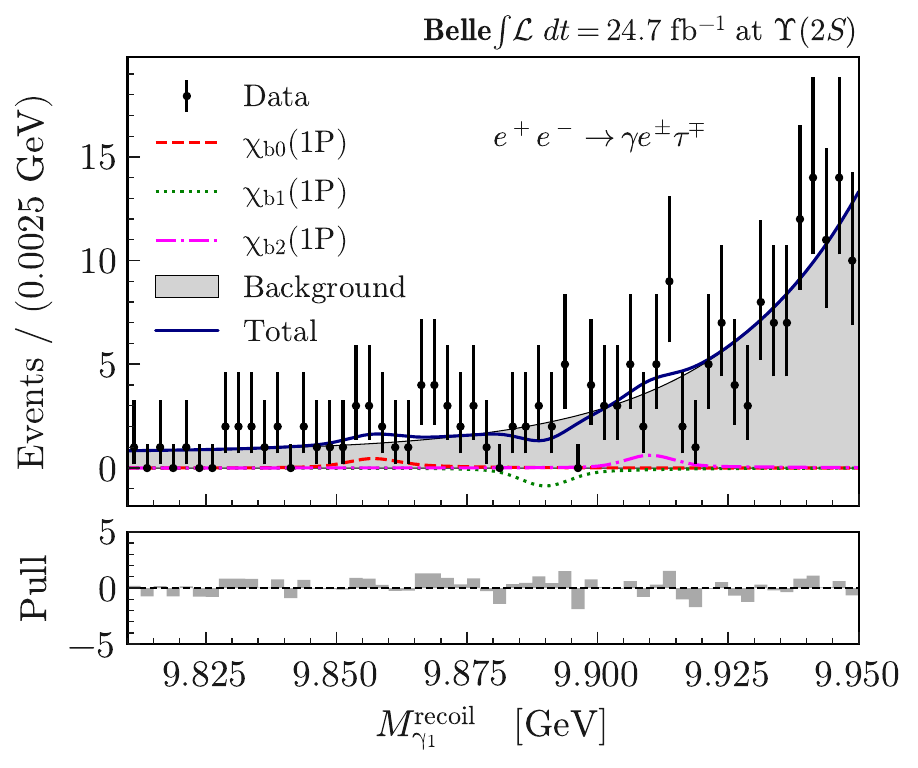}
\includegraphics[width=0.4\textwidth]{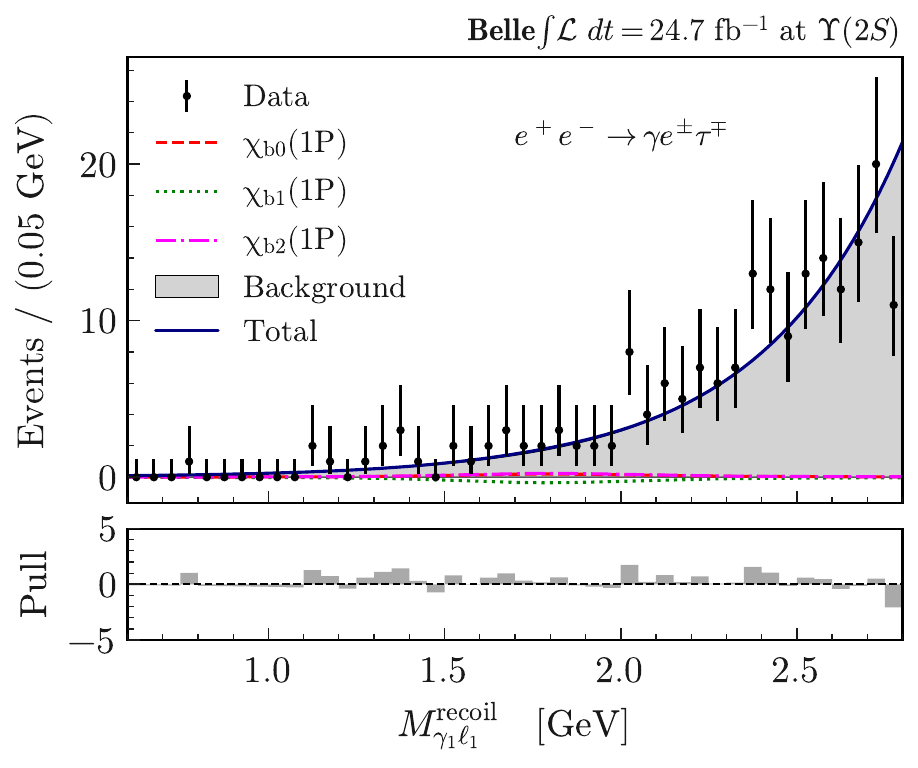}
\includegraphics[width=0.4\textwidth]{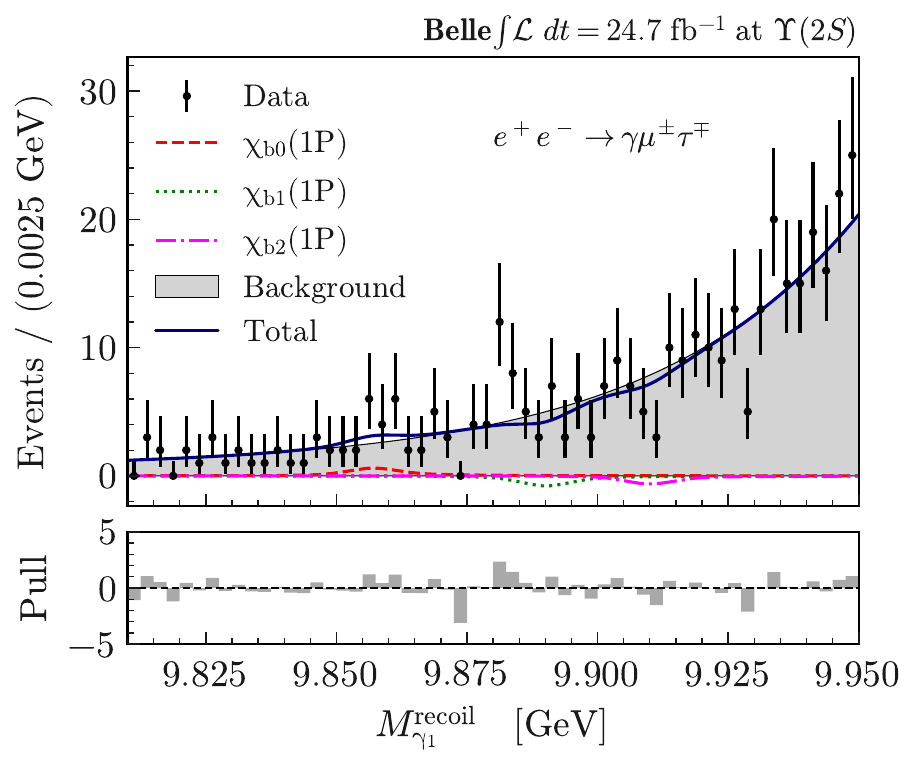}
\includegraphics[width=0.4\textwidth]{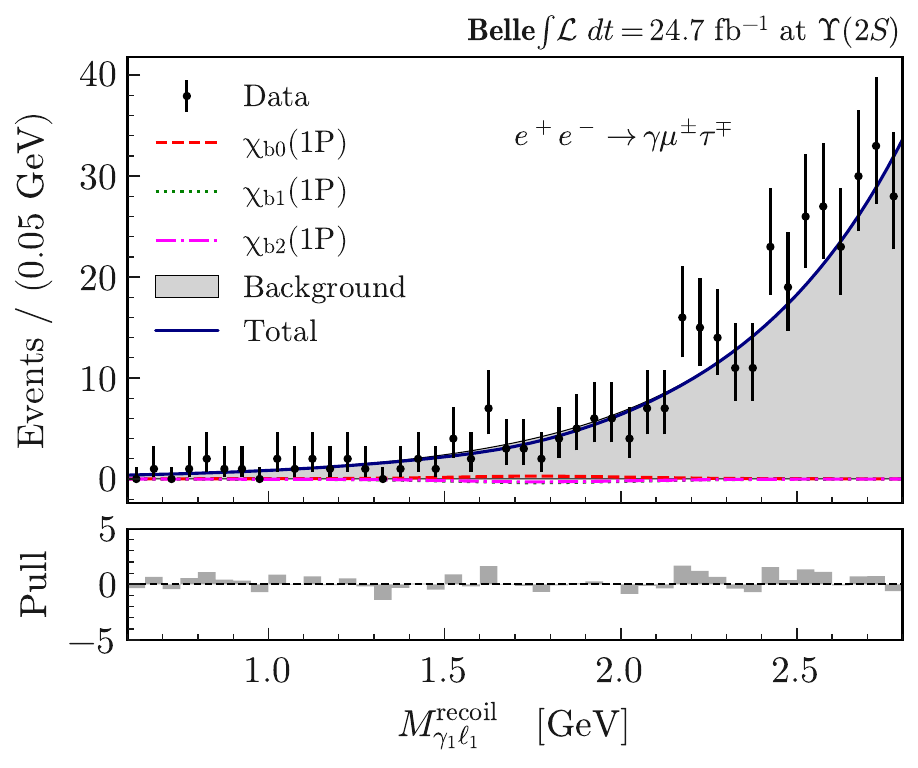}
\end{center}
\caption{Fit projections of $\mrg$ (left) and $\mrgl$ (right) for $\EP\to\GET$ (top row) $\EP\to\GMT$ (bottom row).}
\label{fig:ltau}
\end{figure*}

To extract the signal yields for $\cem{J}$, we perform a maximum-likelihood fit to the $\mrg$ distribution, dividing the signal region into three bins.
Histogram templates for signal components and background are estimated from simulations.
We float the yields of signal components, while the background normalization is constrained by three nuisance parameters describing the systematic uncertainty in each bin. Figure~\ref{fig:em} shows the distribution of $\mrg$ for $\GEM$ candidates in data along with the fitted signal and background components.

For $\clt{J}$, an unbinned maximum-likelihood fit has been performed to the $\mrg$ and $\mrgl$ distributions simultaneously. For both $\mrg$ and $\mrgl$, we use the same signal models as for $\mrg$ in the control modes, with the central values and widths of the signal components fixed to values determined from simulation and corrected by the differences observed between simulation and data in the control modes.
We model the background distribution in $\mrg$ using the background PDF of the control modes and the background distribution in $\mrgl$ with an exponential function. Figure~\ref{fig:ltau} shows the $\mrg$ and $\mrgl$ distributions for $\GET$ and $\GMT$ candidates in data and the results of the fits.

We include systematic uncertainties from the calibration of track-finding, lepton identification, photon detection, and trigger efficiencies, which are corrected in simulation to match the data.
The track-finding systematic uncertainties are 0.35\% per track~\cite{Belle:2012iwr}.
Lepton-identification systematic uncertainties are estimated using control samples of $\jll$ and $\gamma\gamma\to\mu^+\mu^-$~\cite{Hanagaki:2001fz,Abashian:2002bd}.
The photon detection efficiency uncertainties are estimated using radiative Bhabha events.
Trigger systematic uncertainties are estimated to be 1\%, based on the efficiencies of independent sub-triggers for $\EP\to\GGEE$ and $\EP\to\GGMM$ events.
We also include the systematic uncertainties from secondary branching fractions, reconstruction efficiencies, and $\N{Y2}$. Systematic uncertainty arising from the fixed parameters of the PDF are obtained by varying the parameters by $\pm1$ standard deviation in control modes. Uncertainties from the production of $\cb{J}$ are estimated from branching fractions of $\gc{J}$~\cite{ParticleDataGroup:2024cfk} and are the largest systematic uncertainties. Table~\ref{tab:syst} lists all systematic uncertainties and their quadrature sums for each channel.

\begin{table*}[htbp]
  \begin{center}
    \begin{tabular}{l|c|c|c}
      \hline\hline
      \multirow{2}{*}{Source} & \multicolumn{3}{c}{Systematic uncertainties (in \%)} \\ 
      \cline{2-4}
                            & $\EM$              & $\ET$                    & $\MT$                    \\\hline
      $\N{\y2}$               & 2.3              & 2.3                      &2.3                       \\ \hline
      Track finding         & 0.7              & 0.7                      & 0.7                      \\ \hline
      Lepton identification & 1.9, 2.0, 2.0    & 2.2                      & 2.0                      \\ \hline
      Photon detection      & 2.0              & 2.0                      & 2.0                      \\ \hline
      Trigger               & 1.0              & 1.0                      & 1.0                      \\ \hline
      PDF parameters        &                - & $1.1,~0.4,~^{+1.8}_{-1.2}$ & $1.1,~0.4,~^{+1.8}_{-1.2}$ \\ \hline
      Simulation statistics & 0.1              & 0.1                      & 0.1                      \\ \hline
      Secondary branching fractions     & 19.9, 10.0, 10.0 & 19.9, 10.0, 10.0         & 19.9, 10.0, 10.0         \\ \hline
      Total                 & 20.3, 10.7, 10.7 & 20.3, 10.7, 10.8         & 20.3, 10.7, 10.8         \\ \hline\hline
    \end{tabular}
  \end{center}
  \caption{Systematic uncertainties, where three numbers when listed correspond to $J=0, 1,$ and $2$, respectively.}
  \label{tab:syst}
\end{table*}

The branching fraction $\mt{B}(\chi_{bJ}(1P)\to\ell_1\ell_2)$ is obtained from the number of signal events $N_{J, \ell_1\ell_2}$ using
\begin{equation}
\mt{B}(\chi_{bJ}(1P)\to\ell_1\ell_2) = \frac{N_{J, \ell_1\ell_2}/N_{\y2}}{\mt{B}(\gc{J})\epsilon_{J, \ell_1\ell_2}}.
\label{equ:ul}
\end{equation}
Here $\epsilon_{J, \ell_1\ell_2}$ is the efficiency to reconstruct $\gamma\LL$ via the $\chi_{bJ}(1P)$, including the relevant $\tau$ branching fractions.

As all signal yields are consistent with zero, we calculate upper limits on branching fractions using the CLs method~\cite{CLs:2011}.
For $\cem{J}$, we use \pyhf~\cite{Heinrich:2021gyp} and for $\clt{J}$, we use a frequentist calculator from {\tt RooStats}~\cite{Moneta:2010pm}.
In each case, an ensemble of 5000 pseudo-experiments is used.
Systematic uncertainties are included by smearing the yields.
Table~\ref{tab:lfv} lists the efficiencies including the relevant $\tau$ branching fractions, branching fractions, upper limits on them at the 90\% confidence level, as well as the correlation coefficients $C_{mn}$ from simultaneous fits to all three spin states, where $m, n = 0, 1, 2$. For $\cet{J}$, $C_{02}$ is large because of fluctuation in the data.

\begin{table*}[htbp]
  \centering
  \begin{tabular}{l|c|c|c|c|c|c|c|c|c|c|c|c|c|c|c}
  \hline\hline
  \multirow{2}{*}{Decay} &\multicolumn{4}{c|}{$J=0$} &\multicolumn{4}{c|}{$J=1$} &\multicolumn{4}{c|}{$J=2$} & \multicolumn{3}{c}{Correlation}\\
  \cline{2-16}
  &$\epsilon$ &${\rm BR}$ &${\rm UL}$ &${\rm UL^\prime}$ &$\epsilon$ &${\rm BR}$ &${\rm UL}$ &${\rm UL^\prime}$  &$\epsilon$ &${\rm BR}$ &${\rm UL}$ &${\rm UL^\prime}$ &$C_{01}$ &$C_{12}$ &$C_{20}$\\
  \hline
  $\EM$ &28.7 &$1.2\pm1.3$ &$4.0$ &$2.1$  &31.2 &$0.0\pm1.3$ &$1.5$ &1.4  &29.9 & $0.2\pm0.8$  &$1.8$ &1.7  & $-0.01$ & $-4.1$ & $-0.5$\\

  $\ET$ &3.6 &$14.0\pm18.0$ &$41.0$ &18.0 &3.7 &$-13.0\pm8.0$ &$13.0$ &22.0 &3.5 & $9.0\pm12.0$ &$26.0$ &20.0 & $-6.0$  & $5.3$  & $-33.0$\\

  $\MT$ &5.7 &$11.0\pm10.0$ &$28.0$ &32.0 &6.0 &$-8.0\pm5.0$  &$9.0$ &12.0 &5.6 &$-6.0\pm5.0$ &$9.0$ &19.0  &$5.3$  &$8.2$  &$4.2$\\
  \hline\hline
  \end{tabular}
  \caption{Efficiencies ($\epsilon$) in \%, branching fractions (BR) in units of $\e6$, observed  upper limits ($\rm UL$) and expected upper limits ($\rm UL^\prime$) of branching fractions at the 90\% confidence level in units of $\e6$, and correlation coefficients ($C_{mn}$, where $m, n = 0, 1, 2$) from simultaneous fits to all three spin states in \%.}
\label{tab:lfv}
\end{table*}

\begin{figure}[hbtp]
    \begin{center}
    \includegraphics[width=0.9\linewidth]{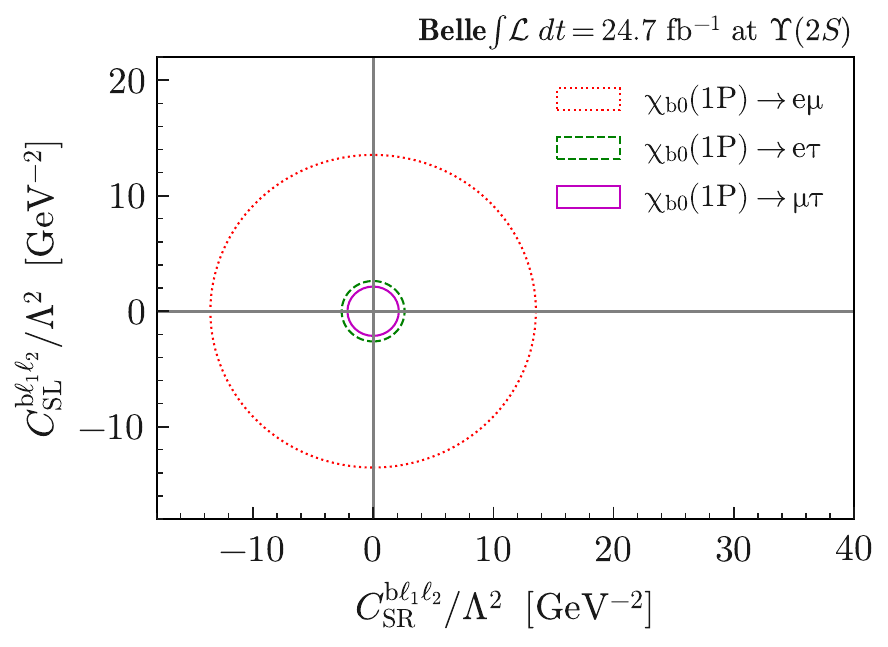}
    \end{center}
    \caption{Upper limits at the 90\% confidence level on Wilson coefficients of scalar operators that mediate CLFV.}
    \label{fig:br_lambda}
  \end{figure}

The Wilson coefficients for the left- and right-handed scalar operators $C_{\rm SL}^{\rm q \ell_1 \ell_2}$ and $C_{\rm SR}^{\rm q \ell_1 \ell_2}$ that mediate CLFV
are constrained by the branching fractions for $\cb0$ decays, relative to the energy scale $\Lambda$ of the new interaction~\cite{Hazard:2016fnc},
using the following equation:
\begin{equation}
  \bigg( \frac {C_{\rm{SL}}^{{\rm{q}}{\ell_1\ell_2}}} {\Lambda^2} \bigg)^2 + \bigg( \frac {C_{\rm{SR}}^{{\rm{q}}{\ell_1\ell_2}}} {\Lambda^2} \bigg)^2 
  = \frac{16\pi m_{\chi} \Gamma_{\chi} {\cal{B}}(\chi \to \ell_1 \ell_2)}{G_{\rm{F}}^2 f_{\chi}^2 m_{\rm{b}}^2 m_{\ell_2}^2 (m_{\chi}^2 - m_{\ell_2}^2)^2}.
  \label{eq:Wilson}
\end{equation}
Here $G_{\rm F}$ is the Fermi constant, $m_{\rm b}$ is the bottom-quark mass, $m_{\ell_2}$ is the heavier lepton mass, and $m_{\chi}$ is the mass of $\cb0$, which are obtained from Ref.~\cite{ParticleDataGroup:2024cfk}. To estimate the total decay width of the $\cb0$, we divide the predicted partial decay width of $\cgy{0}$~\cite{Godfrey:2015} by its measured branching fraction~\cite{ParticleDataGroup:2024cfk}, yielding $\Gamma_{\chi} = 1.23\pm0.17$\mev. The decay constant $f_{\chi}$ is obtained from Ref.~\cite{Chung:2021efj}. We include the uncertainties on $f_{\chi}$ and $\Gamma_{\chi}$ by smearing the results. We obtain bounds at the 90\% confidence level on these Wilson coefficients from equation~\eqref{eq:Wilson} and the branching-fraction upper limits, as shown by the circles in Figure~\ref{fig:br_lambda}. The regions outside the circles are excluded by the obtained results. 

In summary, we present the first searches for CLFV in all nine possible final states with $\ell_1 \neq \ell_2$ and $\ell_1,\ell_2 = e, \mu, \tau$ in $\chi_{bJ}(1P)$ decays ($J = 0,1,2$), and constrain the Wilson coefficients of previously unexplored scalar operators mediating CLFV.

We thank A. Petrov for insightful discussions. This work, based on data collected using the Belle detector, which was
operated until June 2010, was supported by 
the Ministry of Education, Culture, Sports, Science, and
Technology (MEXT) of Japan, the Japan Society for the 
Promotion of Science (JSPS), and the Tau-Lepton Physics 
Research Center of Nagoya University; 
the Australian Research Council including grants
DP210101900, 
DP210102831, 
DE220100462, 
LE210100098, 
LE230100085; 
Austrian Federal Ministry of Education, Science and Research (FWF) and
FWF Austrian Science Fund No.~P~31361-N36;
National Key R\&D Program of China under Contract No.~2022YFA1601903,
National Natural Science Foundation of China and research grants
No.~11575017,
No.~11761141009, 
No.~11705209, 
No.~11975076, 
No.~12135005, 
No.~12150004, 
No.~12161141008, 
and
No.~12175041, 
and Shandong Provincial Natural Science Foundation Project ZR2022JQ02;
the Czech Science Foundation Grant No. 22-18469S;
Horizon 2020 ERC Advanced Grant No.~884719 and ERC Starting Grant No.~947006 ``InterLeptons'' (European Union);
the Carl Zeiss Foundation, the Deutsche Forschungsgemeinschaft, the
Excellence Cluster Universe, and the VolkswagenStiftung;
the Department of Atomic Energy (Project Identification No. RTI 4002), the Department of Science and Technology of India,
and the UPES (India) SEED finding programs Nos. UPES/R\&D-SEED-INFRA/17052023/01 and UPES/R\&D-SOE/20062022/06; 
the Istituto Nazionale di Fisica Nucleare of Italy; 
National Research Foundation (NRF) of Korea Grants
No.~2021R1-A6A1A-03043957,
No.~2021R1-F1A-1064008,
No.~2022R1-A2C-1003993,
No.~2022R1-A2C-1092335,
No.~RS-2016-NR017151,
No.~RS-2018-NR031074,
No.~RS-2021-NR060129,
No.~RS-2023-00208693,
No.~RS-2024-00354342
and
No.~RS-2025-02219521,
Radiation Science Research Institute,
Foreign Large-Size Research Facility Application Supporting project,
the Global Science Experimental Data Hub Center, the Korea Institute of Science and
Technology Information (K25L2M2C3 ) and KREONET/GLORIAD;
the Polish Ministry of Science and Higher Education and 
the National Science Center;
the Ministry of Science and Higher Education of the Russian Federation
and the HSE University Basic Research Program, Moscow; 
University of Tabuk research grants
S-1440-0321, S-0256-1438, and S-0280-1439 (Saudi Arabia);
the Slovenian Research Agency Grant Nos. J1-50010 and P1-0135;
Ikerbasque, Basque Foundation for Science, and the State Agency for Research
of the Spanish Ministry of Science and Innovation through Grant No. PID2022-136510NB-C33 (Spain);
the Swiss National Science Foundation; 
the Ministry of Education and the National Science and Technology Council of Taiwan;
and the United States Department of Energy and the National Science Foundation.
These acknowledgements are not to be interpreted as an endorsement of any
statement made by any of our institutes, funding agencies, governments, or
their representatives.
We thank the KEKB group for the excellent operation of the
accelerator; the KEK cryogenics group for the efficient
operation of the solenoid; and the KEK computer group and the Pacific Northwest National
Laboratory (PNNL) Environmental Molecular Sciences Laboratory (EMSL)
computing group for strong computing support; and the National
Institute of Informatics, and Science Information NETwork 6 (SINET6) for
valuable network support.

\bibliography{reference}

@article{CLEO:2010xuh,
    author = "Kornicer, M. and others",
    collaboration = "CLEO",
    title = "{Measurements of branching fractions for electromagnetic transitions involving the $\chi_{bJ}(1P)$ states}",
    eprint = "1012.0589",
    archivePrefix = "arXiv",
    primaryClass = "hep-ex",
    reportNumber = "CLNS-10-2071, CLEO-10-08",
    doi = "10.1103/PhysRevD.83.054003",
    journal = "Phys. Rev. D",
    volume = "83",
    pages = "054003",
    year = "2011"
}

@article{BaBar:2014och,
    author = "Lees, J. P. and others",
    collaboration = "BaBar",
    title = "{Bottomonium spectroscopy and radiative transitions involving the $\chi_bJ(1P,2P)$ states at BABAR}",
    eprint = "1410.3902",
    archivePrefix = "arXiv",
    primaryClass = "hep-ex",
    reportNumber = "BABAR-PUB-14-005, SLAC-PUB-16123",
    doi = "10.1103/PhysRevD.90.112010",
    journal = "Phys. Rev. D",
    volume = "90",
    number = "11",
    pages = "112010",
    year = "2014"
}

@article{ATLAS:2023mvd,
    author = "Aad, Georges and others",
    collaboration = "ATLAS",
    title = "{Searches for lepton-flavour-violating decays of the Higgs boson into $e\tau$ and $\mu\tau$ in $\sqrt{s}=13$ TeV $pp$ collisions with the ATLAS detector}",
    eprint = "2302.05225",
    archivePrefix = "arXiv",
    primaryClass = "hep-ex",
    reportNumber = "CERN-EP-2022-279",
    doi = "10.1007/JHEP07(2023)166",
    journal = "JHEP",
    volume = "07",
    pages = "166",
    year = "2023"
}

@article{CMS:2021rsq,
    author = "Sirunyan, Albert M and others",
    collaboration = "CMS",
    title = "{Search for lepton-flavor violating decays of the Higgs boson in the $\mu\tau$ and e$\tau$ final states in proton-proton collisions at $\sqrt{s}$ = 13 TeV}",
    eprint = "2105.03007",
    archivePrefix = "arXiv",
    primaryClass = "hep-ex",
    reportNumber = "CMS-HIG-20-009, CERN-EP-2021-061",
    doi = "10.1103/PhysRevD.104.032013",
    journal = "Phys. Rev. D",
    volume = "104",
    number = "3",
    pages = "032013",
    year = "2021"
}

@article{Belle:2021ysv,
    author = "Uno, K. and others",
    collaboration = "Belle",
    title = "{Search for lepton-flavor-violating tau-lepton decays to $\ell\gamma$ at Belle}",
    eprint = "2103.12994",
    archivePrefix = "arXiv",
    primaryClass = "hep-ex",
    doi = "10.1007/JHEP10(2021)019",
    journal = "JHEP",
    volume = "10",
    pages = "19",
    year = "2021"
}

@article{BaBar:2009hkt,
    author = "Aubert, Bernard and others",
    collaboration = "BaBar",
    title = "{Searches for Lepton Flavor Violation in the Decays $\tau^\pm \to e^\pm \gamma$ and $\tau^\pm \to \mu^\pm \gamma$}",
    eprint = "0908.2381",
    archivePrefix = "arXiv",
    primaryClass = "hep-ex",
    reportNumber = "SLAC-PUB-13753, BABAR-PUB-09-026",
    doi = "10.1103/PhysRevLett.104.021802",
    journal = "Phys. Rev. Lett.",
    volume = "104",
    pages = "021802",
    year = "2010"
}

@article{deGouvea:2013zba,
    author = "de Gouvea, Andre and Vogel, Petr",
    title = "{Lepton Flavor and Number Conservation, and Physics Beyond the Standard Model}",
    eprint = "1303.4097",
    archivePrefix = "arXiv",
    primaryClass = "hep-ph",
    doi = "https://doi.org/10.1016/j.ppnp.2013.03.006",
    journal = "Prog. Part. Nucl. Phys.",
    volume = "71",
    pages = "75--92",
    year = "2013"
}

@article{Raidal:2008jk,
    author = "Raidal, M. and others",
    editor = "Fleischer, R. and Hurth, T. and Mangano, M. L.",
    title = "{Flavour physics of leptons and dipole moments}",
    eprint = "0801.1826",
    archivePrefix = "arXiv",
    primaryClass = "hep-ph",
    doi = "https://doi.org/10.1140/epjc/s10052-008-0715-2",
    journal = "Eur. Phys. J. C",
    volume = "57",
    pages = "13--182",
    year = "2008"
}

@article{Celis:2014asa,
    author = "Celis, Alejandro and Cirigliano, Vincenzo and Passemar, Emilie",
    title = "{Model-discriminating power of lepton flavor violating $\tau$ decays}",
    eprint = "1403.5781",
    archivePrefix = "arXiv",
    primaryClass = "hep-ph",
    reportNumber = "FTUV-14-0321, IFIC-14-20, LA-UR-14-21868",
    doi = "https://doi.org/10.1103/PhysRevD.89.095014",
    journal = "Phys. Rev. D",
    volume = "89",
    number = "9",
    pages = "095014",
    year = "2014"
}

@article{Alonso:2015sja,
    author = "Alonso, Rodrigo and Grinstein, Benjam\'\i{}n and Martin Camalich, Jorge",
    title = "{Lepton universality violation and lepton flavor conservation in $B$-meson decays}",
    eprint = "1505.05164",
    archivePrefix = "arXiv",
    primaryClass = "hep-ph",
    doi = "https://doi.org/10.1007/JHEP10(2015)184",
    journal = "JHEP",
    volume = "10",
    pages = "184",
    year = "2015"
}

@article{Abada:2015zea,
    author = "Abada, Asmaa and Be\v{c}irevi\'c, Damir and Lucente, Michele and Sumensari, Olcyr",
    title = "{Lepton flavor violating decays of vector quarkonia and of the $Z$ boson}",
    eprint = "1503.04159",
    archivePrefix = "arXiv",
    primaryClass = "hep-ph",
    reportNumber = "LPT-15-10, SISSA-09-2015-FISI",
    doi = "https://doi.org/10.1103/PhysRevD.91.113013",
    journal = "Phys. Rev. D",
    volume = "91",
    number = "11",
    pages = "113013",
    year = "2015"
}

@article{Hazard:2016fnc,
    author = "Hazard, Derek E. and Petrov, Alexey A.",
    title = "{Lepton flavor violating quarkonium decays}",
    eprint = "1607.00815",
    archivePrefix = "arXiv",
    primaryClass = "hep-ph",
    reportNumber = "WSU-HEP-1603, SI-HEP-2016-19",
    doi = "https://doi.org/10.1103/PhysRevD.94.074023",
    journal = "Phys. Rev. D",
    volume = "94",
    number = "7",
    pages = "074023",
    year = "2016"
}

@article{Godfrey:2015vda,
    author = "Godfrey, Stephen and Logan, Heather E.",
    title = "{Probe of new light Higgs bosons from bottomonium $\chi_{b0}$ decay}",
    eprint = "1510.04659",
    archivePrefix = "arXiv",
    primaryClass = "hep-ph",
    doi = "https://doi.org/10.1103/PhysRevD.93.055014",
    journal = "Phys. Rev. D",
    volume = "93",
    number = "5",
    pages = "055014",
    year = "2016"
}

@article{BaBar:2010vxb,
    author = "Lees, J. P. and others",
    collaboration = "BaBar",
    title = "{Search for Charged Lepton Flavor Violation in Narrow Upsilon Decays}",
    eprint = "1001.1883",
    archivePrefix = "arXiv",
    primaryClass = "hep-ex",
    reportNumber = "SLAC-PUB-13898, BABAR-PUB-09-032",
    doi = "https://doi.org/10.1103/PhysRevLett.104.151802",
    journal = "Phys. Rev. Lett.",
    volume = "104",
    pages = "151802",
    year = "2010"
}

@article{BESIII:2021slj,
    author = "Ablikim, Medina and others",
    collaboration = "BESIII",
    title = "{Search for the charged lepton flavor violating decay $J/\psi\to e\tau$}",
    eprint = "2103.11540",
    archivePrefix = "arXiv",
    primaryClass = "hep-ex",
    doi = "https://doi.org/10.1103/PhysRevD.103.112007",
    journal = "Phys. Rev. D",
    volume = "103",
    number = "11",
    pages = "112007",
    year = "2021"
}

@article{BaBar:2021loj,
    author = "Lees, J. P. and others",
    collaboration = "BaBar",
    title = "{Search for Lepton Flavor Violation in~$\Upsilon (3S)\rightarrow e^{\pm}\mu^{\mp}$}",
    eprint = "2109.03364",
    archivePrefix = "arXiv",
    primaryClass = "hep-ex",
    reportNumber = "BABAR-PUB-21/003, SLAC-PUB-17617",
    doi = "https://doi.org/10.1103/PhysRevLett.128.091804",
    journal = "Phys. Rev. Lett.",
    volume = "128",
    number = "9",
    pages = "091804",
    year = "2022"
}

@article{Belle:2022cce,
    author = "Patra, S. and others",
    collaboration = "Belle",
    title = "{Search for charged lepton flavor violating decays of \Upsilon{} (1S)}",
    eprint = "2201.09620",
    archivePrefix = "arXiv",
    primaryClass = "hep-ex",
    reportNumber = "Belle Preprint 2022-02; KEK Preprint 2021-59",
    doi = "https://doi.org/10.1007/JHEP05(2022)095",
    journal = "JHEP",
    volume = "2022",
    number = "05",
    pages = "095",
    year = "2022"
}

@article{Dhamija:2024,
    author = "Dhamija, R. and others",
    collaboration = "Belle",
    title = "{Search for charged-lepton flavor violation in $\Upsilon(2S)\to\ell^\mp\tau^\pm~(\ell=e, \mu)$ decays at Belle}",
    eprint = "2309.02739",
    archivePrefix = "arXiv",
    primaryClass = "hep-ex",
    reportNumber = "Belle Preprint 2023-14, KEK Preprint 2023-19",
    doi = "https://doi.org/10.1007/JHEP02(2024)187",
    journal = "JHEP",
    volume = "02",
    pages = "187",
    year = "2024"
}

@article{NA62:2021zxl,
    author = "Cortina Gil, Eduardo and others",
    collaboration = "NA62",
    title = "{Search for Lepton Number and Flavor Violation in $K^+$ and $\pi^0$ Decays}",
    eprint = "2105.06759",
    archivePrefix = "arXiv",
    primaryClass = "hep-ex",
    reportNumber = "CERN-EP-2021-090",
    doi = "https://doi.org/10.1103/PhysRevLett.127.131802",
    journal = "Phys. Rev. Lett.",
    volume = "127",
    number = "13",
    pages = "131802",
    year = "2021"
}

@article{White:1995jc,
    author = "White, D. B. and others",
    title = "{Search for the decays eta ---\ensuremath{>} mu e and eta ---\ensuremath{>} e+ e-}",
    reportNumber = "PRINT-95-237 (UCLA)",
    doi = "https://doi.org/10.1103/PhysRevD.53.6658",
    journal = "Phys. Rev. D",
    volume = "53",
    pages = "6658--6661",
    year = "1996"
}

@article{Kurokawa:2001nw,
    author = "Kurokawa, S. and Kikutani, Eiji",
    title = "{Overview of the KEKB accelerators}",
    reportNumber = "KEK-PREPRINT-2001-157A",
    doi = "https://doi.org/10.1016/S0168-9002(02)01771-0",
    journal = "Nucl. Instrum. Meth. A",
    volume = "499",
    pages = "1--7",
    year = "2003"
}

@article{Abe:2013fma,
    author = "Abe, Tetsuo and others",
    title = "Achievements of KEKB",
    doi = "https://doi.org/10.1093/ptep/pts102",
    journal = "PTEP",
    volume = "2013",
    pages = "03A001",
    year = "2013",
}

@article{Belle:2000cnh,
    author = "Abashian, A. and others",
    collaboration = "Belle",
    title = "{The Belle Detector}",
    reportNumber = "KEK-PROGRESS-REPORT-2000-4",
    doi = "https://doi.org/10.1016/S0168-9002(01)02013-7",
    journal = "Nucl. Instrum. Meth. A",
    volume = "479",
    pages = "117--232",
    year = "2002"
}

@article{Belle:2012iwr,
    author = "Brodzicka, Jolanta and others",
    collaboration = "Belle",
    title = "{Physics Achievements from the Belle Experiment}",
    eprint = "1212.5342",
    archivePrefix = "arXiv",
    primaryClass = "hep-ex",
    reportNumber = "KEK-REPORT-2012-5",
    doi = "https://doi.org/10.1093/ptep/pts072",
    journal = "PTEP",
    volume = "2012",
    pages = "04D001",
    year = "2012"
}

@article{Lange:2001uf,
    author = "Lange, D. J.",
    editor = "Erhan, S. and Schlein, P. and Rozen, Y.",
    title = "{The EvtGen particle decay simulation package}",
    doi = "https://doi.org/10.1016/S0168-9002(01)00089-4",
    journal = "Nucl. Instrum. Meth. A",
    volume = "462",
    pages = "152--155",
    year = "2001"
}

@article{Barberio:1993qi,
    author = "Barberio, Elisabetta and Was, Zbigniew",
    title = "{PHOTOS: A Universal Monte Carlo for QED radiative corrections. Version 2.0}",
    reportNumber = "CERN-TH-7033-93",
    doi = "https://doi.org/10.1016/0010-4655(94)90074-4",
    journal = "Comput. Phys. Commun.",
    volume = "79",
    pages = "291--308",
    year = "1994"
}

@article{Brun:1987ma,
    author = "Brun, R. and Bruyant, F. and Maire, M. and McPherson, A. C. and Zanarini, P.",
    title = "{GEANT3}",
    reportNumber = "CERN-DD-EE-84-1",
    note = "{CERN Report No. DD/EE/84-1}",
    month = "9",
    year = "1987"
}

@article{Abashian:2002bd,
    author = "Abashian, A. and others",
    title = "{Muon identification in the Belle experiment at KEKB}",
    doi = "https://doi.org/10.1016/S0168-9002(02)01164-6",
    journal = "Nucl. Instrum. Meth. A",
    volume = "491",
    number = "1-2",
    pages = "69--82",
    year = "2002"
}

@article{Hanagaki:2001fz,
    author = "Hanagaki, K. and Kakuno, H. and Ikeda, H. and Iijima, T. and Tsukamoto, T.",
    title = "{Electron identification in Belle}",
    eprint = "hep-ex/0108044",
    archivePrefix = "arXiv",
    doi = "https://doi.org/10.1016/S0168-9002(01)02113-1",
    journal = "Nucl. Instrum. Meth. A",
    volume = "485",
    pages = "490--503",
    year = "2002"
}

@article{Belle-II:2023izd,
    author = "Adachi, I. and others",
    collaboration = "Belle II",
    title = "{Measurement of the \ensuremath{\tau}-lepton mass with the Belle II experiment}",
    eprint = "2305.19116",
    archivePrefix = "arXiv",
    primaryClass = "hep-ex",
    reportNumber = "Belle II Preprint 2023-008, KEK Preprint 2023-6",
    doi = "https://doi.org/10.1103/PhysRevD.108.032006",
    journal = "Phys. Rev. D",
    volume = "108",
    number = "3",
    pages = "032006",
    year = "2023"
}

@article{ParticleDataGroup:2024cfk,
    author = "Navas, S. and others",
    collaboration = "Particle Data Group",
    title = "{Review of particle physics}",
    doi = "https://doi.org/10.1103/PhysRevD.110.030001",
    journal = "Phys. Rev. D",
    volume = "110",
    number = "3",
    pages = "030001",
    year = "2024"
}

@article{Heinrich:2021gyp,
    author = "Heinrich, Lukas and Feickert, Matthew and Stark, Giordon and Cranmer, Kyle",
    title = "{pyhf: pure-Python implementation of HistFactory statistical models}",
    doi = "https://doi.org/10.21105/joss.02823",
    journal = "J. Open Source Softw.",
    volume = "6",
    number = "58",
    pages = "2823",
    year = "2021"
}

@article{Wang:2011ny2,
  title = {Search for charmonium and charmoniumlike states in $\ensuremath{\Upsilon}(2S)$ radiative decays},
  author = {Wang, X. L. and others},
  collaboration = {Belle},
  journal = {Phys. Rev. D},
  volume = {84},
  issue = {7},
  pages = {071107},
  numpages = {7},
  year = {2011},
  month = {Oct},
  eprint = "1008.1774",
  archivePrefix = "arXiv",
  publisher = {American Physical Society},
  doi = {https://doi.org/10.1103/PhysRevD.84.071107},
  url = {https://link.aps.org/doi/10.1103/PhysRevD.84.071107}
}

@article{Sjöstrand_2006,
doi = {10.1088/1126-6708/2006/05/026},
url = {https://doi.org/10.1088/1126-6708/2006/05/026},
year = {2006},
month = {may},
publisher = {},
volume = {2006},
number = {05},
pages = {026},
author = {Torbjörn Sjöstrand and Stephen Mrenna and Peter Skands},
title = {PYTHIA 6.4 physics and manual},
journal = {JHEP},
}

@article{CLs:2011,
title = {Asymptotic formulae for likelihood-based tests of new physics},
journal = {Eur. Phys. J. C},
volume = {71},
pages = {1554},
year = {2011},
eprint = {1007.1727},
archivePrefix = {arXiv},
primaryClass = {physics.data-an},
doi = {https://doi.org/10.1140/epjc/s10052-011-1554-0},
author = {G. Cowan and K. Cranmer and E. Gross and O. Vitells},
}

@article{Moneta:2010pm,
    author = "L. Moneta and others",
    title = "{The RooStats Project}",
    journal = "PoS",
    volume = "ACAT2010",
    pages = "057",
    year = "2010",
    eprint = "1009.1003",
    archivePrefix = "arXiv",
    primaryClass = "physics.data-an",
    doi = "https://doi.org/10.22323/1.093.0057"
}

@article{CHEON2002548,
title = {Electromagnetic calorimeter trigger at Belle},
journal = {Nucl. Instrum. Meth. A},
volume = {494},
pages = {548-554},
year = {2002},
doi = {https://doi.org/10.1016/S0168-9002(02)01547-4},
url = {https://www.sciencedirect.com/science/article/pii/S0168900202015474},
author = {B.G Cheon and others},
}

@article{Chung:2021efj,
    author = "Chung, Hee Sok",
    title = "{P-wave quarkonium wavefunctions at the origin in the $ \overline{\mathrm{MS}} $ scheme}",
    eprint = "2106.15514",
    archivePrefix = "arXiv",
    primaryClass = "hep-ph",
    reportNumber = "TUM-EFT 145/21",
    doi = "10.1007/JHEP09(2021)195",
    journal = "JHEP",
    volume = "09",
    pages = "195",
    year = "2021"
}

@article{Godfrey:2015,
  title = {Bottomonium mesons and strategies for their observation},
  author = {Godfrey, Stephen and Moats, Kenneth},
  journal = {Phys. Rev. D},
  volume = {92},
  issue = {5},
  pages = {054034},
  numpages = {39},
  year = {2015},
  month = {Sep},
  doi = {10.1103/PhysRevD.92.054034}
}
\bibliographystyle{apsrev4-1}

\end{document}